\documentclass[letter]{aa}

\usepackage{graphicx}
\usepackage{txfonts}
\begin{document}

   \title{Declining metallicity and extended \ion{He}{II} in the outflow of an epoch of reionization analogue galaxy}

   \subtitle{}

   \author{M. J. Hamel-Bravo\inst{1}
          \fnmsep \thanks{email: mhamelbravo@swin.edu.au}
          \and
          D. B. Fisher\inst{1}
          \and
          D. A. Berg\inst{2}
          \and
          A. J. Cameron\inst{3}
          \and
          J. Chisholm\inst{2}
          \and
          G. G. Kacprzak\inst{1}
          \and
          B. Mazzilli Ciraulo\inst{1}
          \and 
          H. Katz\inst{4}
          }

   \institute{Centre for Astrophysics and Supercomputing, Swinburne University of Technology, Hawthorn, Victoria 3122, Australia
          \and
             Department of Astronomy, University of Texas, Austin, TX 78712, USA
          \and
             Sub-department of Astrophysics, University of Oxford, Keble Road, Oxford OX1 3RH, United Kingdom
          \and
             Department of Astronomy \& Astrophysics, University of Chicago, 5640 S Ellis Avenue, Chicago, IL 606337, USA
          }

   \date{Received September 15, 1996; accepted March 16, 1997}

 
  \abstract
   {We present VLT/X-shooter spectroscopy of the extremely metal-poor starburst galaxy SBS~0335‑052E, a nearby (D$\sim$54~Mpc) analog of high redshift systems, probing its outflow up to a distance of $\sim$2.6~kpc. Using direct-method oxygen abundances, we find a complex metallicity profile that generally declines with distance, decreasing by 0.37~dex from the galaxy center out to 2.4~kpc into the outflow. This implies a metal-loading factor roughly an order of magnitude lower than predictions based on the mass–metallicity relation for low-mass galaxies. We also detect extended \ion{He}{II} emission, including a broad, redshifted component beyond 2~kpc, distinct from narrow emission associated with star clusters. No H$\beta$ emission is detected associated with the broad redshifted component, implying extremely high \ion{He}{II}/H$\beta$ ratios (0.7 - 4.8). Such extreme values, combined with the decreasing metallicity, challenge our current models for stellar feedback. The contribution of an IMBH could simultaneously account for the declining metallicity and the unusually extended \ion{He}{II} emission.}
   





   \maketitle
%

\section{Introduction}

Galactic outflows are a critical component of galaxy evolution. They are frequently invoked in simulations as drivers of the mass–metallicity relationship (MZR) \citep[e.g.][]{Finlator2008} and the metal enrichment of the circumgalactic medium (CGM) \citep[e.g.][]{Christensen2018, Peroux2020}, by transporting metal-rich supernovae ejecta from the interstellar medium (ISM), into the surrounding circumgalactic medium. 


Both simulations and analytical models require outflows in metal-poor, low-mass galaxies to have relatively larger ratios of metallicity in the outflow compared to the galaxy than higher mass systems \citep{Muratov2015, Sharda2024}. This enrichment is commonly expressed through the metal-loading factor ($\zeta$), which quantifies the ratio of metals carried away by the outflow to the metals produced by stars.

The detection of metals in the intergalactic medium (IGM) at $z > 4$ \citep[e.g.][]{DOdorico2013} points to efficient early enrichment, likely via galactic outflows. JWST spectra reveal many similarities between galaxies in the Epoch of Reionization with local Universe low metallicity starburst galaxies \citep{Schaerer2022}. The metal-loading is a critical component of early Universe mass evolution, but would be impractical to observe directly at $z>5$, due to the faintness of the gas. Nearby analogues are, therefore, a critical tool to study the physical mechanisms at play in the baryon cycle during the early Universe. 

Observational constraints on metal-loading are rare because they require a self-consistent approach to compare the metallicity of the ISM (Z$_{\rm ISM}$) to that of the outflow (Z$_{\rm out}$) \citep[e.g.][]{Chisholm2018,Cameron2021,Hamel-Bravo2024}.  Recent studies have successfully characterized Z$_{\rm ISM}$ and Z$_{\rm out}$ in nearby galaxies using the auroral-line method \citep{Cameron2021, Hamel-Bravo2024}. Moreover, \cite{Hamel-Bravo2025} show, by comparing different species of auroral lines, that this method appears robust in galactic winds.  


SBS~0335-052E is a nearby ($D \sim 54$ Mpc), low-mass ($M_\star \sim 10^7$–$10^8 M_\odot$), extremely metal-poor ($Z \sim 1/40 Z_\odot$), highly-ionized galaxy. It is often described as a high-redshift analogue \citep[e.g.][]{Izotov1997,Kehrig2018,Mingozzi2025}. It hosts an outflow extending $\geq 15$~kpc from the galaxy \citep{Herenz2023}. This work presents the metallicity profile along the outflow of SBS~0335-052E using auroral lines reaching $\sim$2.4~kpc from the brightest region of the galaxy. 

\section{Data}

   \begin{figure*}
   \centering
   \includegraphics[width=1\textwidth]{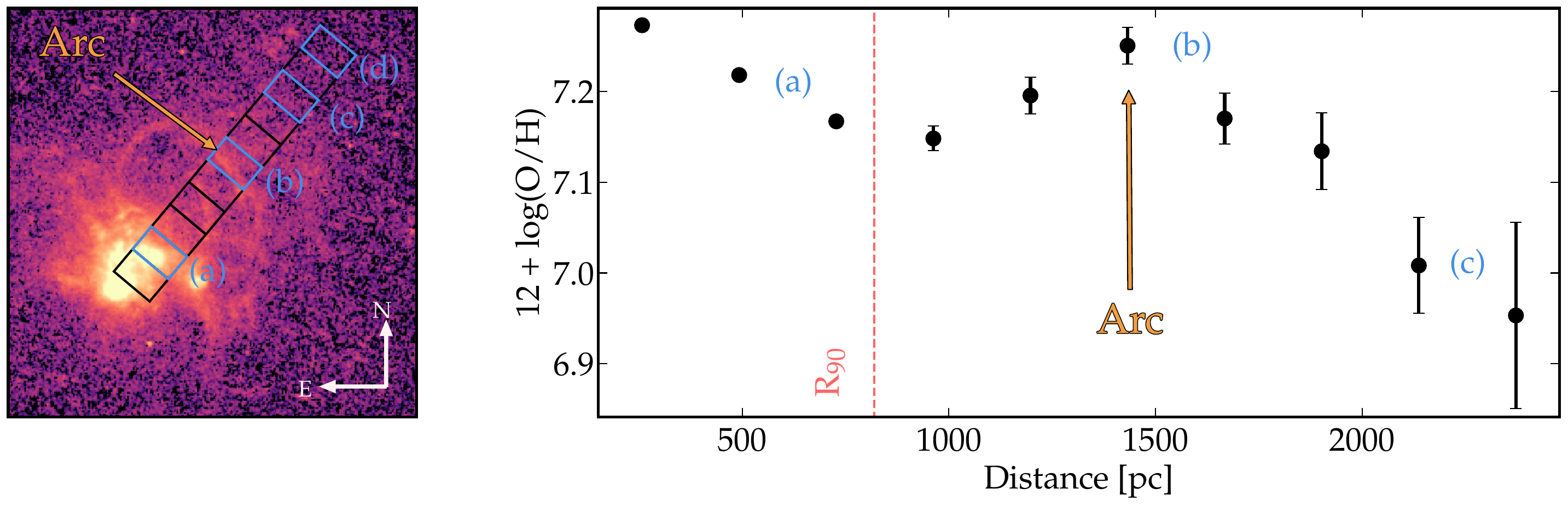}

   \caption{Left panel: X-shooter slit position overlaid on an HST/WFC continuum-subtracted H$\alpha$ image. We show the 1\arcsec division of the slit and mark relevant positions with blue letters. Right panel: The metallicity profile as a function of distance to the brightest H$\alpha$ emission in the galaxy. The red dotted line shows the R$_{90}$ measured from the HST F550n band. Note the rise in metallicity corresponds to the H$\alpha$ emission arc.}
              \label{fig_1}%
    \end{figure*}
    
SBS~0335-052E was observed with X-shooter on the Very Large Telescope (VLT) (Program ID: 112.25V8.001). The slit was placed at a position angle of -30$^{\circ}$ aiming at the extended ionized gas, NW of the galaxy \citep{Herenz2023}. The left panel of Fig.~\ref{fig_1} shows the slit position overlaid on a continuum-subtracted HST/WFC H$\alpha$ image (Program ID: 10575). These data were presented in detail by \cite{Hamel-Bravo2025}; here we provide a summary of the observations and reduction.  

Observations were carried out with seeing conditions of 0.6-1.1\arcsec. The slit is 11\arcsec $\times$ $\sim$1.6\arcsec. The median spectral resolution is $FWHM\sim75$~km\,s$^{-1}$. The total exposure time was 7380~s for the UVB arm and 7228~s for the VIS arm, divided into 5 exposures. Dedicated sky exposures were obtained separately. We used ESO Recipe Flexible Execution Workbench (Reflex) \citep{Freudling2013} for the data reduction. The slit was divided into 11 segments of 1\arcsec each, resulting in 11 spectra (Fig.~\ref{fig_1}).

We corrected each spectrum for Milky Way foreground extinction using the \cite{Cardelli1989} extinction law with an A$_V$ = 0.127 (NASA/IPAC Extragalactic Database), and then removed the stellar continuum. For the three spatial elements closest to the galaxy (continuum SNR~>~5), we fit the continuum with the Penalized PiXel-Fitting (pPXF; \cite{Cappellari2017}) and GALAXEV models \citep{Bruzual2003}. At larger distances, where the continuum SNR was too low, we fit a 2$^{nd}$ order polynomial. We correct for internal extinction using Balmer line ratios (see \cite{Hamel-Bravo2025}). 

We measure emission-line fluxes by fitting Gaussian models with constrained centroids, minimum $\sigma$ set to the average instrumental dispersion, 0.5~$\AA$, and height and width as free parameters. In H$\alpha$ and \ion{He}{II}~$\lambda$4686, the non-Gaussian shape indicates the presence of multiple components, so we use 2-Gaussian models. Not all spatial elements support 2-Gaussians, we therefore use the Bayesian Information Criterion (BIC), with $\Delta$~BIC$\sim$100 threshold \citep[as discussed in][]{ReichardtChu2025}, to assess the need for multi-Gaussian models. Flux uncertainties were derived from 100 Monte Carlo simulations, where each spectrum was perturbed by its pixel errors, and the standard deviation of the recovered fluxes is adopted as the flux uncertainty.

\section{Declining metallicity profile}

We measure oxygen abundance from the [\ion{O}{III}]$\lambda$4363 auroal line in all X-shooter positions except the one farthest from the galaxy (position (d) in Fig.~\ref{fig_1}). We calculate the electron density ($n_e$) from the [\ion{O}{II}]$\lambda\lambda$3727,3729 ratio and $T_e$ from the [\ion{O}{III}]$\lambda$4363/5007 ratio. For $T_e$([\ion{O}{II}]), we adopt the $T_e$–$T_e$ relation from \cite{Campbell1986}. Ionic abundances of O$^+$ and O$^{++}$ are derived using the corresponding $T_e$. We assume negligible contributions from higher ionization states. To estimate uncertainties, we use the emission line flux uncertainties. 


The right panel of Fig.~\ref{fig_1} shows the metallicity profile as a function of distance to the brightest H$\alpha$ emission. The metallicity at the closest position to the galaxy is $12+\log(\rm O/H)=7.273\pm0.002$, consistent with the value found by \cite{Papaderos2006}. The metallicity decreases with distance until reaching a $12+\log(\rm O/H)=7.14\pm0.01$ at $\sim$1~kpc. It then rises again to a local maximum of $12+\log(\rm O/H)=7.25\pm0.02$ at the position of the arc, at $\sim$1.5~kpc. Beyond the arc, the metallicity profile decreases steadily, reaching a minimum of $12+\log(\rm O/H)=6.9\pm0.1$ at our maximum probed distance of $\sim$2.4~kpc. 

The ionized gas morphology of SBS~0335-052E exhibits multiple filaments and arc-like structures aligned with our X-shooter slit \citep{Thuan1997, Herenz2017}. Such arcs trace feedback-driven superbubbles, where hot gas fills the cavity and cooler gas forms swept-up shells that can break out of the ISM \citep{MacLow1988}. The left panel of Fig.~\ref{fig_1} shows a bright H$\alpha$ arc at $\sim$1.5~kpc from the galaxy in the NW direction. At the position of the arc, we require two Gaussians to fit the H$\alpha$ emission. The best fit shows a bright H$\alpha$ component blueshifted from the systemic velocity by $\sim$80~km\,s$^{-1}$. The arc exhibits an inhomogeneous structure, and large ionized filaments are observed extending beyond it. \cite{Herenz2017} interpreted this as the bubble having entered its ``blow-out'' phase, during which a large-scale outflow develops and ionizing radiation, and possibly metals, can escape into the CGM.

The metallicity profile of SBS~0335-052E shows substructure that changes at similar locations to the H$\alpha$ emission. We find a metallicity peak at the arc and a local minimum in the interior. Similar metal-enriched shells surrounded by low-metallicity gas were reported in I~Zw~18 \citep{RickardsVaught2025}. This likely reflects incomplete mixing of supernova ejecta with the warm ionized medium, with compressed ejecta at the bubble edge producing the observed peak. If the 80~km\,s$^{-1}$ blueshifted H$\alpha$ emission traces the velocity of the enriched outflow, then the projected distance from the brightest H$\alpha$ knot to the arc (1.4~kpc) implies a travel time of $\sim$18~Myr, during which mixing has not yet occurred. 


Simulations of outflows predict a systematic increase in the warm ionized gas phase metallicity with distance, with metal-poor galaxies having a steeper increase \citep{Kim2020, Vijayan2024}. Observationally, \cite{Cameron2021} and \cite{Hamel-Bravo2024} found that the outflow metallicity profiles of Mrk~1468 (Z$\sim 0.1~Z_{\odot}$) and NGC~1569 (Z$\sim 0.3~Z_{\odot}$) qualitatively follow the trends predicted by simulations. SBS~0335-052E, unlike previous observations, deviates from standard expectations and simulations. This suggests that current models may not fully capture the processes of metal transport in extremely metal-poor galaxies and that these could be different in such extreme systems.

   \begin{figure}
   \centering
   \includegraphics[width=0.49\textwidth]{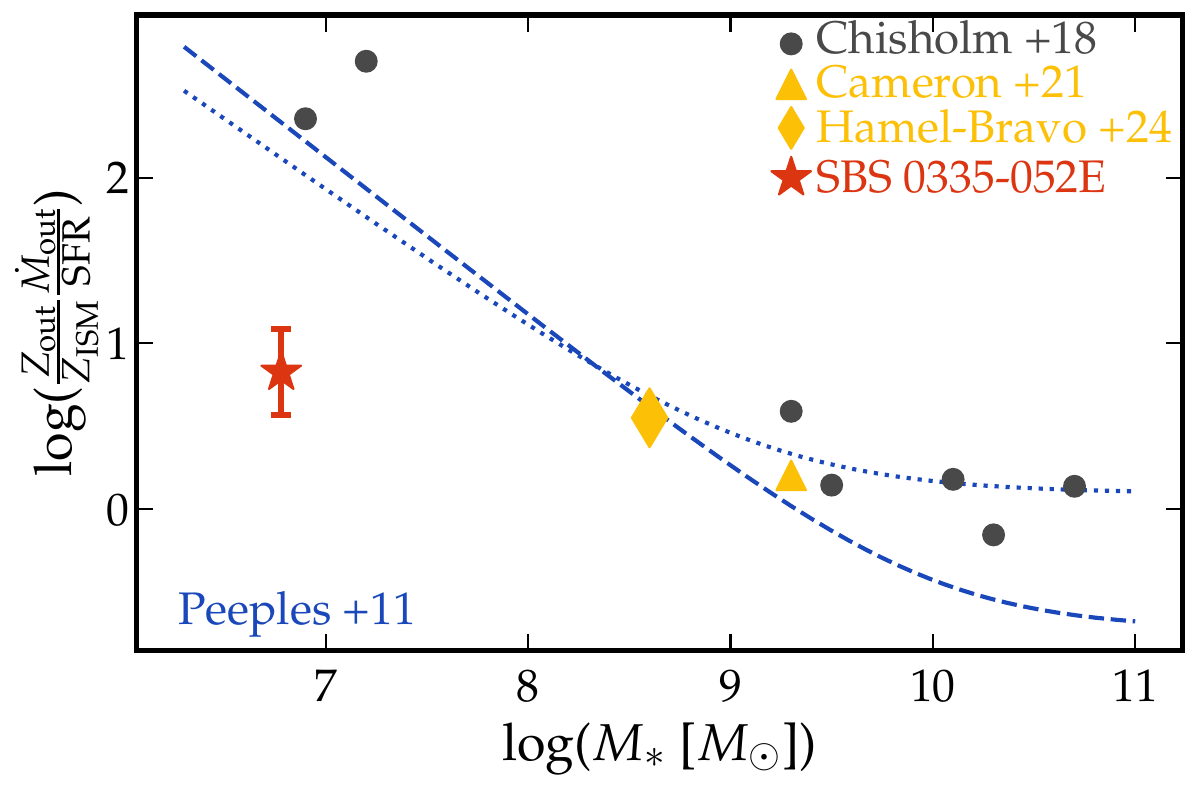}
   \caption{Metal loading vs. stellar mass. The red star shows SBS~0335-052E, black dots show results from \protect\cite{Chisholm2018}, and the yellow triangle and diamond show results from \protect\cite{Cameron2021} and \protect\cite{Hamel-Bravo2024}, respectively. Blue dotted and dashed lines show models from \protect\cite{Peeples2011} to reproduce the MZR from \protect\cite{Denicol2002} and \protect\cite{Tremonti2004}, respectively.}
              \label{fig:metal-loading}%
   \end{figure}
   
We estimate the metal-loading factor, defined as $\zeta = (Z_{out}/Z_{ISM})(\dot M_{out}/\mathrm{SFR})$. We separate the ISM and outflow regions in the metallicity profile using the radius that encloses 90\% (R$_{90}$) of the stellar light in the HST/WFC F550M image (red line in Fig.~\ref{fig_1}). Based on a 15~kpc conical outflow model, \citet{Herenz2023} estimated a mass-loading factor of $\dot M_{out}/\mathrm{SFR} \geq 10$ for the ionized gas. Assuming our 2.7~kpc metallicity profile represents the full outflow, and taking the mean metallicity inside $R_{90}$ as $Z_{ISM}$=7.24 and beyond $R_{90}$ as $Z_{out}$=7.13, we calculate $\zeta = 5.75$. Upper and lower limits for $\zeta$ are calculated by assuming the lowest and highest metallicity measured beyond $R_{90}$, respectively.

Fig.~\ref{fig:metal-loading} shows $\zeta$ as a function of stellar mass. SBS~0335-052E  (red star) is shown along with previous observations of absorption line studies and auroral line measurements. Model predictions from \cite{Peeples2011}, calibrated to reproduce the different MZR, are shown for comparison.

Our measurement is more than an order of magnitude below model predictions and absorption line studies at this stellar mass. A higher mass-loading of $\sim$100 could reconcile this, but this would be much higher than typical values for dwarf galaxies (0.1 – 20) \citep[e.g.][]{Chisholm2017, McQuinn2019}. Since our assumed value of 10 is similar to those values used in the absorption line study shown in Fig.~\ref{fig:metal-loading}, the difference most likely arises from $Z_{out}/Z_{ISM}$. This may reflect the difference in gas phases traced, diffuse gas along the line of sight versus denser extended gas. Systematic offsets between different metallicity measurement methods may also contribute to this discrepancy \citep{Kewley2008}. However, such offsets are not observed for higher-mass galaxies, and together with the contrasting metallicity profile we measure, this points to a different enrichment process operating in SBS~0335-052E.

SBS~0335-052E resembles the metal-poor, low-mass galaxies thought to dominate at high redshift. Its weakly enriched outflow suggests that galaxies in the early universe may not have enriched their surroundings as efficiently as often predicted by models \citep{Oppenheimer2006, Muratov2015}. If this behavior is representative, additional physical processes may be needed to fully explain the enrichment of the IGM at early cosmic times and the MZR. Further observations of extremely metal-poor outflows and higher distances along the outflow will be essential to put more constraints on models and simulations.


%


\section{Extended \ion{He}{II} emission}

   \begin{figure}
   \centering
   \includegraphics[width=0.27\textwidth]{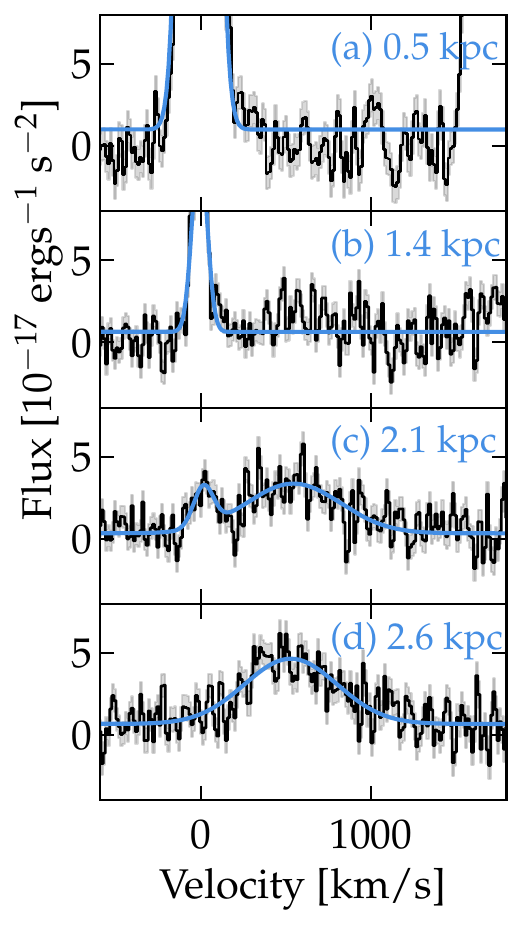}
   \caption{Spectra around the \ion{He}{II}~$\lambda$4686 emission line, in velocity scale, for the four spatial elements marked in blue in Fig.~\ref{fig_1}. v = 0 corresponds to the redshift of the galaxy, z = 0.01352. Black shows the data and blue shows the Gaussian fit.}
              \label{fig:HeII}%
   \end{figure}

We detect the \ion{He}{II}~$\lambda$4686 emission line in the 11 spatial elements along the X-shooter slit. Fig.~\ref{fig:HeII} shows the spectra around the \ion{He}{II} emission for the four positions highlighted in Fig.~\ref{fig_1}. 

For all spectra at distances z~<~1.4~kpc from the brightest H$\alpha$ knot, the \ion{He}{II} emission is well described by a single Gaussian centered at v~$\approx$~0, with a median FWHM of $\sim$129~km\,s$^{-1}$. Beyond the arc, from z~=~1.4~-~2.1~kpc (position c), the emission is best modeled with a narrow component and a broad, redshifted component. At the two positions farthest from the galaxy ($\sim$2.1–2.6~kpc), only the broad, redshifted component is detected. The broad component is redshifted relative to the narrow component by a median of $\sim$507~km\,s$^{-1}$ and has a median FWHM of $\sim$655~km\,s$^{-1}$. The strength of the narrow \ion{He}{II} emission decreases with increasing distance from the galaxy, while a broad redshifted component simultaneously increases.

\cite{Kehrig2018} reported extended \ion{He}{II}$\lambda$4686 emission up to $\sim$1.6kpc to the NW with median FWHM matching the narrow component in our data. From our data, the narrow \ion{He}{II}/H$\beta$ has a peak value of $\sim$0.055 at position a (Fig.\ref{fig_1}), consistent with the bright knot identified by \cite{Kehrig2018} ($\sim$0.06). The ratio declines to $\sim$0.026 at $\sim$2.1~kpc, where we detect the furthest narrow \ion{He}{II}. Such high \ion{He}{II}/H$\beta$ values are difficult to reproduce with stellar population models \citep{Shirazi2012}, indicating additional ionizing sources. Proposed origins of strong \ion{He}{II} and other high ionization emission lines in SBS~0335-052 include metal-free rotating stars or a very low-metallicity binary population with a top-heavy IMF \citep{Kehrig2018}, a luminous blue variable outburst \citep{Peng2025}, an intermediate-mass black hole (IMBH) \citep{Hatano2023}, or a combination of processes \citep{Mingozzi2025}.

All these theories are constructed on observations of high-ionized emission with associated emission in other strong nebular lines. We detect no H$\beta$ or H$\alpha$ emission with the same FWHM or velocity as the broad \ion{He}{II} emission observed beyond 1.4~kpc. Assuming the narrow H$\beta$ emission is unrelated to the broad \ion{He}{II}, we estimate an upper limit for broad H$\beta$ using a 3$\sigma$ detection threshold and adopting the same FWHM as the broad \ion{He}{II}. This corresponds to $F_{\mathrm{H}\beta} \approx 8.7 \times 10^{-18}\mathrm{erg\,s^{-1}\,cm^{-2}}$ at the furthest slit position (position d), which corresponds to a lower limit of \ion{He}{II}/H$\beta$ = 4.8, an exceptionally high value. Alternatively, integrating both lines over the velocity range of the broad \ion{He}{II} gives \ion{He}{II}/H$\beta \sim 0.7$, still elevated compared to typical values. 

Broad \ion{He}{II} emission can arise from  Wolf–Rayet (WR) stars, shocks, and active galactic nuclei (AGN). Although one star cluster in SBS~0335-052E hosts WR stars \citep{Papaderos2006}, the broad \ion{He}{II} emission we detect lies at projected distances $>$2~kpc from the galaxy center, making an origin of stars colocated with \ion{He}{II} unlikely. The \ion{He}{II} luminosity  at position d, $\sim 3.1 \times 10^{37},\mathrm{erg,s^{-1}}$, would require $\sim$90 WR stars \citep[based on values for IZw18;][]{Crowther2006}, yet no stellar continuum is detected. Radiative shock models \citep{Allen2008} predict maximum \ion{He}{II}/H$\beta$ ratios of $\sim$0.3 at SMC-like metallicities. While higher ratios are possible at solar metallicity, this is inconsistent with the extremely low metallicity of SBS~0335‑052, and BPT diagnostics are not consistent with shocks in this region. 

The fast, highly ionized gas traced by the broad redshifted \ion{He}{II} emission, at distances $>$2~kpc, coincides with the decline in metallicity. One possibility is that the oxygen is locked in higher ionization states and is not accounted for in our O/H estimates. However, the \ion{He}{II} emission is redshifted and broader than the nebular oxygen and hydrogen lines, suggesting it originates from a different kinematic component. \cite{Berg2021} showed that in nebular emission lines of extreme emission line galaxies, \ion{O}{IV} contributes 1-2\% of the total O/H. 
Because there are no other lines detected with a similar velocity as \ion{He}{II}, its metallicity is highly uncertain. 
Notably, SBS-0335-052E shows extended X-ray emission along the outflow, which may be associated with the \ion{He}{II} emission. 


\cite{Mingozzi2025} argue that the high ratio of [\ion{Ne}{V}]/[\ion{Ne}{II}] in the star cluster of SBS~0335-052E suggests the presence of a $\sim10^5$~M$_{\odot}$ intermediate mass black hole (IMBH) that contributes to the ionization of the galaxy. While alternative interpretations exist \citep{Peng2025}, feedback from an IMBH could explain both the declining metallicity and the bright \ion{He}{II} emission. The $\sim$500~km/s redshifted, highly ionized gas seen in \ion{He}{II} may trace feedback from the IMBH rather than pure star formation. Such IMBH-driven feedback would not carry the metal-rich supernova ejecta of massive stars, leading to lower metallicity than gas accelerated by stellar feedback.  At large distances from the galaxy we observe both (1) fast-moving, highly ionized gas and (2) decreasing metallicity in the warm ionized phase. While \ion{He}{II} is in general difficult to explain in this galaxy \cite{Kehrig2018}, its existence far from significant stellar sources seems even more difficult to explain. Moreover, the high velocity of the \ion{He}{II} is likewise difficult to explain from traditional expectations of stellar feedback \citep[e.g.][]{Kim2020}.
An additional source of both ionization of the gas and mechanical feedback, such as an IMBH, could be key to understanding the decreasing metallicity in the wind of SBS~0335-052E.



\section{Summary}

Using VLT/X-shooter spectroscopy, we probe the outflow of the extremely metal-poor starburst galaxy SBS~0335-052E to projected distances of $\sim$2.6~kpc. Detection of [\ion{O}{III}]$\lambda$4363 out to $\sim$2.4~kpc enables direct measurements of the metallicity profile, revealing a peak at a bright H$\alpha$ arc followed by a steady decline. This contrasts with simulations that predict metal-rich winds in low-mass galaxies. The computed metal-loading factor is substantially lower than estimates based on the mass–metallicity relation and absorption-line studies. This discrepancy raises essential questions about how metals are retained, redistributed, or distributed among different ionization phases in such systems, with important implications for chemical enrichment at high redshift.

In addition, we detect extended \ion{He}{II}~$\lambda$4686 emission out to $\sim$2.6 kpc. While the inner regions show narrow emission consistent with nebular ionization, at larger distances we observe a secondary broader and redshifted component. This is indicative of another ionizing source responsible for such emission. This broad component is difficult to account for with current models of stellar populations, shocks, or other ionizing mechanisms. It may trace a hotter outflowing phase, which could contain additional unaccounted metals or may be the result of an extra feedback mechanism like an IMBH.


\begin{acknowledgements} 
This draft was improved by discussions with Mark Krumholz. Based on observations collected at the European Organisation for Astronomical Research in the Southern Hemisphere under ESO programme(s) 112.25V8.001.
\end{acknowledgements}

%
%

\bibliographystyle{aa}

\bibliography{SBS_paper_bib.bib}

\end{document}